\documentclass[prb,amsmath,amssymb,twocolumn,superscriptaddress,showpacs]{revtex4-1}
\usepackage{bm}
\usepackage{amssymb}
\usepackage{colordvi}
\usepackage{graphicx}
\usepackage{color}
\usepackage{hyperref}
\newcommand{\up}{\uparrow}
\newcommand{\down}{\downarrow}

\newcommand{\be}{\begin{equation}}
\newcommand{\ee}{\end{equation}}
\newcommand{\bea}{\begin{eqnarray}}
\newcommand{\eea}{\end{eqnarray}}

\newcommand{\vep}{\varepsilon}

\newcommand{\ome}{\omega}

\def\Re {\mbox{Re}}

\def\grad {\mbox{\boldmath$\nabla$\unboldmath}}

\begin{document}

\title{Accidental degeneracy and topological phase transitions in two-dimensional core-shell dielectric photonic crystals}

\author{Lin Xu}\email{These authors contribute equally to this work}
\author{HaiXiao Wang}\email{These authors contribute equally to this work}
\author{YaDong Xu}
\author{HuanYang Chen}\email{chy@suda.edu.cn}
\author{Jian-Hua Jiang}\email{joejhjiang@sina.com}
\affiliation{College of Physics, Optoelectronics and Energy, \&
  Collaborative Innovation Center of Suzhou Nano Science and
  Technology, Soochow University, 1 Shizi Street, Suzhou 215006, China}

\date{\today}

\begin{abstract}
A simple core-shell two-dimensional photonic crystal is studied where
the triangle lattice symmetry and the $C_{6v}$ point group symmetry
give rich physics in accidental touching points of photonic bands. We
systematically evaluate different types of accidental nodal points at
${\vec k}=0$ for transverse-magnetic harmonic modes when the geometry and
permittivity of the core-shell material are continuously tuned. The
accidental nodal points can have different dispersions and topological
properties (i.e., Berry phase). These accidental nodal points can be
the critical states lying between a topological phase and a normal
phase of the photonic crystal. They are thus very important for the
study of topological photonic states. We show that, without breaking
time-reversal symmetry, by tuning the geometry of the core-shell
material, a phase transition into the photonic quantum spin Hall
insulator can be achieved. Here the ``spin'' is defined as the orbital
angular momentum of a photon. We study the topological phase
transition as well as the properties of the edge and bulk states and
their application potentials in optics.
\end{abstract}

\pacs{42.70.Qs,78.67.Pt,03.65.Vf}

\maketitle

\section{Introduction}

The Ginzburg-Landau paradigm\cite{GL} gained wide success in describing phase
transitions before the discovery of the quantum Hall effect\cite{qhe,fqhe}. The
emergence of topological states of matter raises new challenges in
understanding the states of matter and phase transitions. Topological
states of matter offer new physics and applications in electronic
systems such as topological insulators and topological semimetals
(e.g., graphene\cite{graphene}). The conical dispersions in topological semimetal
and the helical edge states of topological insulators are found to be
useful for a lot of applications\cite{TIreview}. Since Haldane and Raghu's seminal
work which brought the quantum anomalous Hall effect to the realm of
photonics\cite{Raghu}, more and more researches are focused on the topological
effects in photonic crystals (PhCs)\cite{phTI}.

PhCs are periodically arranged dielectric materials which break the
continuous translation symmetry that gives the dispersion of light in
homogeneous materials. The resulting photonic spectrum is the photonic
energy bands similar to the electronic energy bands in solid
crystals\cite{eli,sajeev}. The emergence of photonic energy bands and band gaps
introduces tremendous advantages in manipulating light\cite{book}. Unlike
previous studies on trapping light with photonic band gaps, in this
work we focus on the nodal points in photonic bands in two-dimensional
(2D) triangle PhCs. We examine their relationships with Dirac (and
Dirac-like) points and topological phase transitions. Nontrivial Berry
phases assign a topological winding number to the nodal points, making
them mother states of quantum spin Hall insulators and quantum
anomalous Hall insulators (denoted as the $Z_2$ and $Z$ topological
insulators, respectively)\cite{TIreview,qahe}. The topological nodal points studied here
are induced by accidental degeneracy's instead of deterministic
degeneracy's due to, e.g., nonsymmorphic or sublattice symmetry (e.g.,
in photonic  graphene\cite{ph-gra}). Accidental degeneracy in 2D PhCs have
attracted a lot of research attention in recent years. In square
lattice photonic crystals, a conical Dirac-like dispersion\cite{ct2d} was found
theoretically. At the Dirac point the photonic crystal can be regarded
as a lossless effective medium of zero refractive index\cite{ct2d,Sakoda}.

Photonic analog of time-reversal symmetric $Z_2$ topological
insulators using bi-anisotropic metamaterials was
proposed\cite{z2-ph}, where the coupling between the electric field
and magnetic field plays the same role as the spin-orbit coupling in
electronic $Z_2$ topological insulators. Recently, the topological
phase transition between normal photonic band gap materials and a
photonic $Z_2$ topological insulator was proposed in 2D PhCs using
normal dielectric materials with isotropic
permittivity\cite{huxiao}. The 2D PhCs that realize such a transition
is a triangle lattice where there are six cylinders in each unit cell
forming a pattern with $C_{6}$ symmetry. By tuning the position of the
cylinders a photonic band gap carrying $Z_2$ topology can be formed.
At the boundary between the $Z_2$ topological phase and the trivial
phase, a double Dirac cone is formed precisely at the geometry that
the lattice becomes a honeycomb\cite{huxiao}. Therefore, the double
Dirac cone is induced by a deterministic degeneracy due to the
sublattice symmetry of the honeycomb lattice. Recently Dirac-cones are
studied in PhCs\cite{Sakoda,pgs} and phononic crystals\cite{mei}.

In this work, we propose a simpler realization of the photonic analog
of the $Z_2$ topological insulator, using core-shell dielectric
materials. There is only a single core-shell cylinder at the center of
each unit cell. The core-shell structure has
continuous rotation symmetry which is compatible with the largest
point group symmetry $C_{6v}$ of the triangle lattice.
This provides the condition for the realization of the $Z_2$ topological
phase. Moreover, it off the possibilities to achieve various
degeneracy's such as Dirac-like cone, double Dirac cone, and quadratic
band touching. We systematically calculate the accidental degeneracy's
at the Brillouin zone center and the topological phase transitions in
the 2D core-shell dielectric PhCs.  The core-shell triangle photonic
crystal structure is simple and mechanically stable and hence it is
compatible with colloidal\cite{colloid} self-assembled structure as
well as biological systems\cite{bio}. Our study shows how topological
phases of photons can be realized in simple 2D dielectric PhCs and
various phases can be induced by tuning the geometry of the dielectric
materials.

\begin{figure}
\begin{center}
\includegraphics[width=8.6cm]{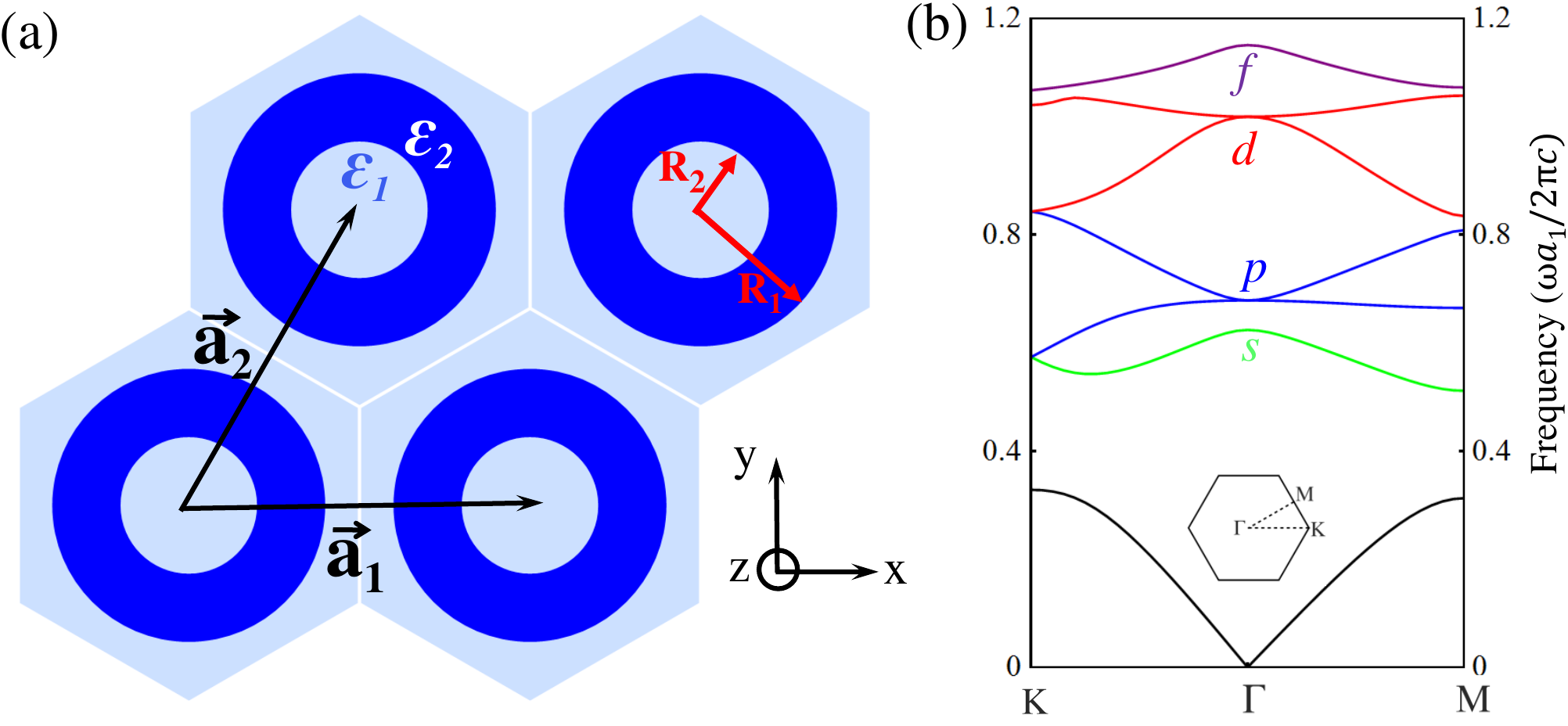}
\caption{ (Color online) (a) Schematic configuration of a triangle PhC
  using core-shell dielectric materials. The dielectric constant of
  the core-shell dielectric material and background material are
  $\vep_2$ and $\vep_1$, respectively. $R_1$ and $R_2$ are the outer
  and inner radii of the core-shell cylinder. $\vec{a}_1$ and
  $\vec{a}_2$ are the two basis vectors of the triangle lattice. We
  set $a_1=a_2\equiv 1$ as the lattice constant. The inverse
  structure is defined by exchange of the permittivity $\vep_1$ and
  $\vep_2$. (b) A typical band structure of the core-shell triangle
  lattice PhC with $R_1=0.16a_1$, $R_2=0.05a_1$, $\vep_1=1$ and
  $\vep_2=12$. We use the symbols $s, p, d$ and $f$ to label the modes
  at the $\Gamma$ point. 
}
\label{fig1}
\end{center}
\end{figure}

\section{2D TRIANGLE PHOTONIC CRYSTALS}
Among common 2D PhCs, triangle photonic crystals have the largest
point group symmetry\cite{pgs}. As will be shown below the 2D triangle PhCs can
host several kinds of topological nodal points. These nodal points are
the mother states of different types of photonic topological
insulators. In the triangle lattice, the $\Gamma$ point has two kinds
of 2D irreducible representations in the $C_{6v}$ symmetry, which are
denoted as the $E_1$ and $E_2$ representations,
respectively\cite{sa-book}. The associated Bloch waves are referred as the
$E_1$ and $E_2$ modes, respectively. Each representation has two Bloch
waves, which can be reorganized into a pair of modes connected by
time-reversal operation. Here the relevant modes are the
$p_{+}=p_x+ip_y$ and $p_-=p_x-ip_y$ modes, which are denoted as
(pseudo-) spin up and spin down states of the $p$ bands. Similarly, the
$d_{+}=d_{x^2-y^2}+id_{xy}$ and $d_-=d_{x^2-y^2}-id_{xy}$ modes are
denoted as the (pseudo-) spin up and spin down states for the $d$
bands. 

The 2D core-shell dielectric structure in triangle lattice is shown in
Fig.~1(a). Its outer and inner radii are denoted as $R_1$ and $R_2$,
respectively. The dielectric constant of the core-shell material and
the background material are $\vep_1$ and $\vep_2$, respectively. The
lattice constant is set as $a_1=a_2\equiv 1$. We define the inverse
structure by the 
exchange of $\vep_1$ and $\vep_2$. The inverse structure is hence a
core-shell air cylinder in dielectric background. Our architecture is
one of the simplest 2D PhCs that can support rich physics of
topological nodal points and topological phase transitions. 

By numerically solving the Maxwell's equations for transverse
magnetic (TM) harmonic modes, of which the electric field is along the
core-shell cylinder (i.e., the $z$ direction). Fig.~1(b) shows a typical
band structure of the PhC. The band structure is calculated by COMSOL
based on the finite-element method. In this band structure, the
eigen-modes at the $\Gamma$ point exhibit orbital symmetry of the $s,
p, d$ and $f$ waves from low frequency to high frequency. These modes
correspond to the $A_1$, $E_1$, $E_2$ and $B_1$ modes of the $C_{6v}$
point group symmetry\cite{sa-book}. The labeling of bands with $s$,
$p$, $d$ and $f$ modes is only effective at the $\Gamma$ point which
possesses the $C_{6v}$ symmetry. Away from the $\Gamma$ point these
modes mix with each other and their order in frequency may
change. Nevertheless, the labels clearly reveal the accidental band
degeneracy and band inversion at the $\Gamma$ point. Besides, the $s,
p, d$ and $f$ modes carry physical meanings. The photonic bands can be
viewed as derived from transfer (hopping) of local Mie resonances of
the core-shell structures between adjacent unit cells. Such a
tight-binding understanding of the photonic bands has successfully
connected the photonic bands with the Mie resonances of $s, p$ and $d$
symmetries\cite{tb,kivshar}. Thus the Mie resonances can be regarded
as the ``atomic orbits'' for photonic energy bands, as it plays the
same role as the atomic orbits in electronic energy bands.

\begin{figure}
\begin{center}
\includegraphics[width=8.7cm]{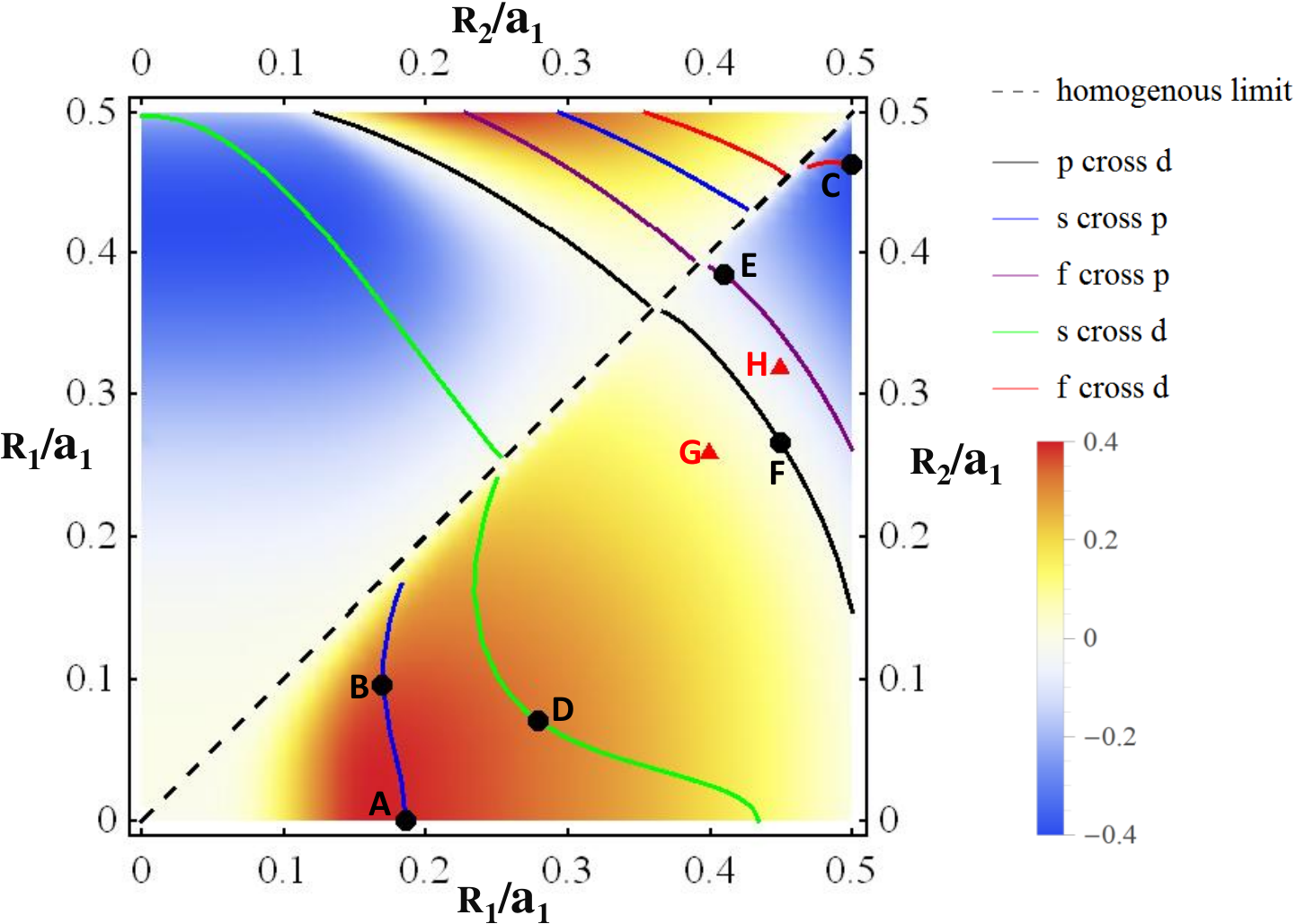}
\caption{ (Color online) Phase diagram of the $p$-$d$ inversion induced
  photonic $Z_2$ topological insulator in the $R_1$-$R_2$ parameter
  space. The diagonal line represents the homogeneous limit which
  separates the normal and reversed structures in the upper and lower
  triangular regions. The lower region is the parameter space for
  core-shell dielectric cylinder in air background (called as the
  normal structure), while the upper region is the parameter space for
  core-shell air cylinder in dielectric background. The dielectric
  constant for air and dielectric are $\vep_1=1$ and $\vep_2=12$,
  respectively. The contour color represents the value of
  $\Delta\ome_{pd}$. The red region represents that the $p$ bands 
  are lower than the $d$ bands at the $\Gamma$ point, whereas the blue region
  stands for the photonic analog of the $Z_2$ topological insulators with
  the $p$ bands above the $d$ bands at the $\Gamma$ point. The $p$-$d$ bands
  degeneracy at the $\Gamma$ point is labeled by the black line. The $s$-$p$,
  $p$-$f$, $s$-$d$, $f$-$d$ band inversions are also calculated and
  plotted. Several points in the phase diagram are labeled, which
  will be used for the discussions in the main text.
}
\label{fig2}
\end{center}
\end{figure}

The $s$ and $f$ ($A_1$ and $B_1$) modes are singlet states, while the
$E_1$ and $E_2$ modes are doubly degenerate, according to the $C_{6v}$
symmetry. By continuously changing the radii $R_1$ and $R_2$ or the
dielectric constant, different bands cross each other, leading to
accidental degeneracy's at the $\Gamma$ point. The accidental
degeneracy usually results in linear or quadratic dispersion. The
scale invariance of the Maxwell’s equations dictates that the
independent variables in our system are the inner and outer radii
(divided by the lattice constant) as well as the ratio of the two
dielectric constants $\vep_1/\vep_2$. The properties of the photonic 
bands for $R_2=0$ are well-studied in the literature, where no
topological nodal point or topological band structure is found.

\section{PHASE DIAGRAM}

We first discuss the eigen-frequency of the $p$ and $d$ bands at the
$\Gamma$ point. We use the following dimensionless quantity to
characterize the relative $p$-$d$ band-gap size at the $\Gamma$ point, 
\be
\Delta\ome_{pd} = 2\frac{\ome_d-\ome_p}{\ome_d+\ome_p} ,
\ee
where $\ome_p$ and $\ome_d$ are the egien-frequency's of the $p$ and
$d$ modes at the $\Gamma$ point, respectively.

By numerically calculating the photonic band structure, we obtain  in
the $R_1$-$R_2$ parameter space with $\vep_1=1$ and $\vep_2=12$ (see
Fig.~2). The figure clearly demonstrates the $p$-$d$ band inversion
which can result in photonic analog of the $Z_2$ topological
insulator. The properties of this photonic $Z_2$ topological states
will be discussed in details below. We incorporate in the upper
triangle of Fig.~2 the phase diagram for the reversed structure, where
the core-shell cylinder of air induces the photonic energy bands. The
diagonal line is the homogeneous limit, which can be regarded as no
dielectric or no air (for the inverse structure). The influence of
$\vep_1/\vep_2$ on the phase diagram will be discussed below. 

The color scheme represents: the red region has $p$ bands lower (in
frequency) than the $d$ bands at the $\Gamma$ point, while the blue
region has $p$ bands higher than the $d$ bands at the $\Gamma$ point
(i.e., $p$-$d$ band inversion). The blue region thus represents the
$Z_2$ topological phase ($Z_2=1$), while the red region represents the
normal phase (topologically trivial, $Z_2=0$). The parity inversion at
the $\Gamma$ point for the $p$-$d$ reversed photonic bands dictate the
$Z_2$ topology\cite{fuprb}. At the boundary between
the topological phase and the normal phase, the double Dirac-cone
dispersion emerges at the $\Gamma$ point. The properties of the
photonic $Z_2$ topological insulator will be discussed in details
below.

In addition, we use the blue (red) curve in Fig.~2 to denote the
crossing between the $s$ and $p$ bands (the $f$ and $d$ bands) in the
phase diagram. These two situations lead to Dirac-like cones at the
$\Gamma$ point. The crossing between the $s$ and $d$ bands (the $f$
and $p$ bands) are denoted as the green and purple curves in the phase
diagram. These crossings lead to quadratic band touching at the
$\Gamma$ point. All these accidental band degeneracy's and the
properties of the nodal point will be discussed in details below.

Other important information revealed by Fig.~2 is the geometric
conditions for the emergence of various nodal points in the photonic
spectrum. These nodal points are important in several reasons. First,
they are the mother states of topologically nontrivial states of
photons. An example has been demonstrated above for the relation
between the double Dirac cone and the photonic $Z_2$ topological
insulator. The Berry phase of each band is closely related to the
formation of the photonic topological states. Usually, topological
nodal points are the critical states between normal band gaps and
topological insulators. By introducing time/parity breaking
perturbations, the double-Dirac-cone state can become a photonic
quantum anomalous Hall insulator ($Z$ topological insulator) or a
$Z_2$ topological insulator, or a trivial photonic band gap material,
depending on the specific gap opening perturbations\cite{onoda}. 
Our study includes the triangle photonic crystals with dielectric rods
as a special limit of $R_2$=0. From the phase diagram, it is clear that
the $p$-$d$ band inversion cannot be attained using dielectric rods with
$R_2$=0. Thus the dielectric rod photonic crystals cannot support double
Dirac cone. They can only support Dirac-like cone due to $s$-$p$
degeneracy or quadratic band touching due to $s$-$d$ degeneracy. For the
inverse structure a simple air cylinder can only support $s$-$d$
degeneracy.

We plot in Fig.~3 the band structures of different kinds of accidental
degeneracy's marked by the black points in Fig.~2. In Figs.~3(a) and
3(b), the linear dispersion due to the accidental degeneracy of the
singlet $s$ mode and the doubly degenerate $p$ modes are plotted, leading
to the Dirac-like cone. The Dirac-like cone is associated with a cone
and a flat band intersecting at the Dirac point. Its effective
Hamiltonian is ${\cal H}=v_0{\vec k}\cdot {\vec S}$  where $\vec{S}$
is the pseudo-spin 1 consisting of three modes $s$, $p_x$, and
$p_y$\cite{ct2d}. $v_0$ is the group velocity around the Dirac
point. The upper, lower and flat bands have pseudo-spin along the wave
vector direction as 1, -1, and 0, respectively.

The Dirac-like cone can emerge in photonic crystals with $R_2=0$
(cylinder rods) as well as for finite $R_2$ (hollow cylinders). The
photonic dispersions for these two situations are illustrated in
Figs.~3(a) and 3(b), respectively. Another Dirac-like cone dispersion
is shown in Fig.~3(c) due to accidental degeneracy of the singlet $f$
mode and the doubly degenerate $d$ modes. Linear, conical dispersion
appears for the $s$-$p$ and $d$-$f$ degeneracy's because the parity of
the $s$ and $p$ ($d$ and $f$) modes are different. According to the
${\vec k}\cdot{\vec P}$ theory\cite{mei2,mei3}, the linear in $k$ coupling
between two photonic bands originates from the $P$ matrix element
which is finite only between bands with different parities. For bands
with the same parity, such coupling is quadratic\cite{mei2,mei3}. The
quadratic dispersion around the touching point of the $s$ and $d$
bands is shown in Fig.~3(d), while the quadratic band touching around
the $f$-$p$ degeneracy is shown in Fig.~3(e).

The accidental degeneracy of the $p$ doublets and the $d$ doublets
[Fig.~3(f)] carries nontrivial topological properties which are the
main focus of this work. Since the parity of the $p$ modes and the $d$
modes are different, the dispersion around the accidental degeneracy is
conical. The two fold degeneracy's of the $p$ and $d$ bands result in a
double Dirac cone that resembles the dispersion of Dirac's famous
equation for electron and positron with vanishing mass in 2D systems.

\begin{widetext}

\begin{figure}
\begin{center}
\includegraphics[width=13cm]{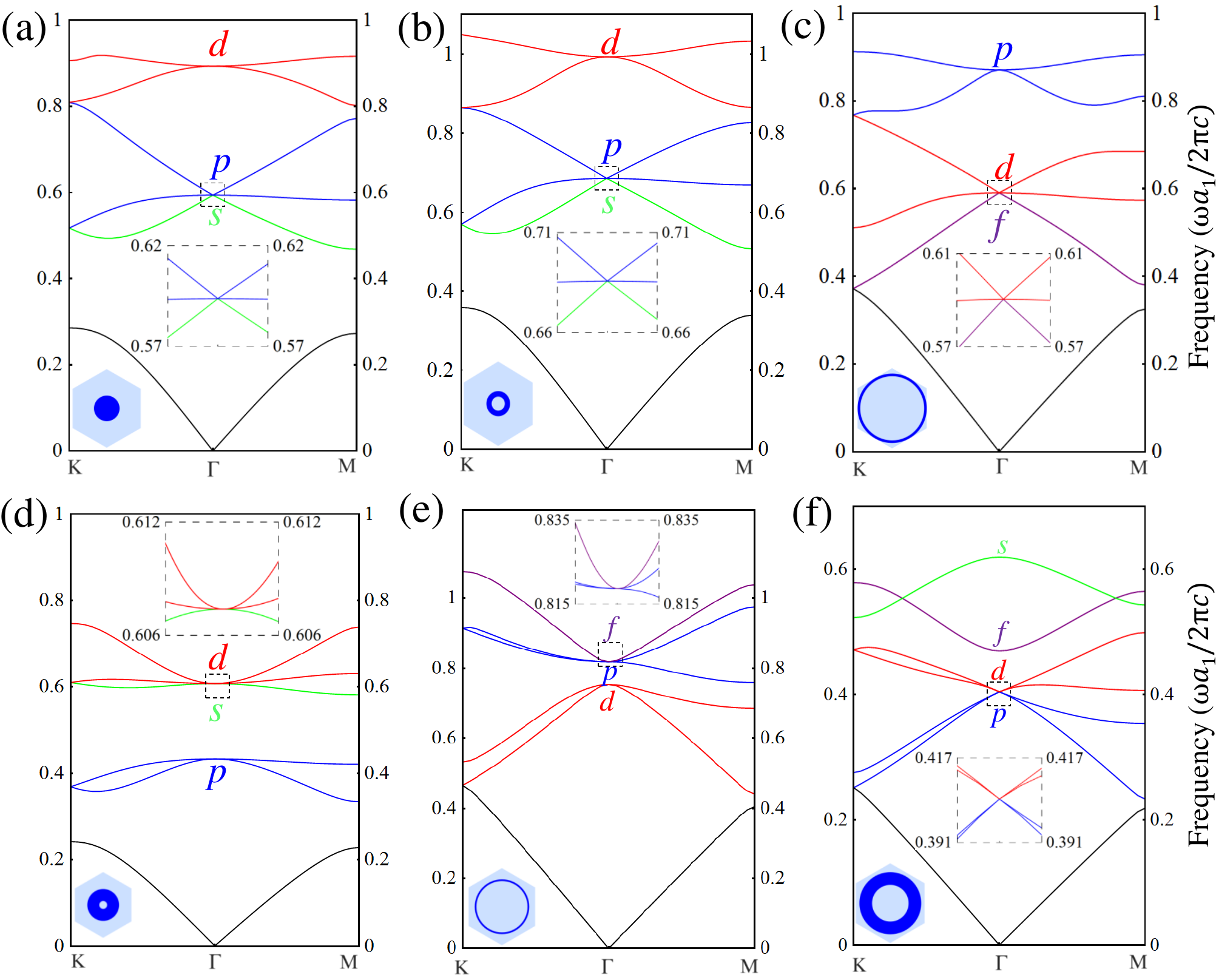}
\caption{ (Color online) Band structures of different kinds of
  accidental degeneracy. (a) The band structure of the $A$ point in the
  phase diagram with $R_1=0.1816$ and $R_2=0$. (b) The band structure of the
  $B$ point in the phase diagram with $R_1=0.17$ and $R_2=0.0947$. (c) The
  band structure of the $C$ point in the phase diagram with $R_1=0.5$ and
  $R_2=0.4624$. (d) The band structure of the $D$ point in phase diagram
  with $R_1=0.28$ and $R_2=0.0697$. (e) The band structure of the $E$ point in
  the phase diagram with $R_1=0.41$ and $R_2=0.3844$. (f) The band structure
  of the $F$ point in the phase diagram with $R_1=0.45$ and
  $R_2=0.2656$. The geometry of the core-shell material in a unit cell
  is shown at the left-down corner of each figure. The 
  colored curves labeled with $s, p, d$, and $f$ represent the modes at the
  $\Gamma$ point only (instead of the whole bands). We zoom in the dispersion near
  the nodal points with a dashed frame. }
\label{fig3}
\end{center}
\end{figure}

We now exploit the ${\vec k}\cdot{\vec P}$ theory to derive the
effective Hamiltonian for the photonic bands near the nodal
points. The Maxwell's equation for the TM modes can be written as 
\be
\grad\times \frac{1}{\vep({\vec r})} \grad\times {\vec
  h}_{n,\vec{k}}({\vec r}) = \frac{\ome^2}{c^2} {\vec
  h}_{n,\vec{k}}({\vec r}) ,
\ee
where $n$ is the band index and ${\vec h}_{n,\vec{k}}({\vec r})$ is
the Bloch function of the magnetic field of photon. The Bloch function
is normalized as $\int_{u.c.} d{\vec r} {\vec
  h}_{n^\prime,\vec{k}}^\ast({\vec r}){\vec h}_{n,\vec{k}}({\vec
  r})=\delta_{nn^\prime}$ with $u.c.$ denoting the unit cell (i.e.,
integration in a unit cell). The Hermitian operator $\grad\times
\frac{1}{\vep({\vec r})} \grad\times$ can be viewed as the photonic
Hamiltonian. Expanding the Bloch function ${\vec h}_{n,\vec{k}}({\vec
  r})$  in the basis of the Bloch wavefunctions at the $\Gamma$ point,
${\vec h}_{n,0}({\vec r})$, one can establish a ${\vec k}\cdot{\vec
  P}$ Hamiltonian,
\be
H_{nn^\prime} ({\vec k}) = \frac{\ome_{n,0}^2}{c^2}\delta_{nn^\prime} +
{\vec k}\cdot {\vec P}_{nn^\prime}  - \int_{u.c.} 
\frac{d{\vec r}}{\vep({\vec r})} \vec{h}_{n,0}^\ast({\vec
  r})\cdot[\vec{k}\times(\vec{k}\times \vec{h}_{n^\prime,0}({\vec r}))] ,
\ee
where $\ome_{n,0}$ is the eigen-frequency of the $n^{th}$ band at the
$\Gamma$ point. The matrix element of ${\vec P}$ is given by 
\be
{\vec P}_{nn^\prime} = \int_{u.c.}\frac{d\vec{r}}{\vep({\vec r})} [\vec{h}_{n^\prime,0}^\ast({\vec
  r})\times(i\grad\times \vec{h}_{n,0}^\ast({\vec r}))+(i\grad\times
\vec{h}_{n^\prime,0}({\vec r}))\times \vec{h}_{n,0}^\ast({\vec
  r})]  .
\ee
We notice that the matrix element of ${\vec P}$ is nonzero only when
the $n$ and $n^\prime$ bands are of different parity. Using the above
${\vec k}\cdot{\vec P}$ theory, to the linear order in $\vec{k}$, the
effective Hamiltonian of the $p$ and $d$ bands is written in the basis
of $(p_+,p_-,d_+,d_-)^T$ as, 
\be
{\cal H} = \left(\begin{array}{cccc} \frac{\ome_p^2}{c^2} & 0 & A k_+ & 0  \\
  0 & \frac{\ome_p^2}{c^2} & 0 & A^\ast k_- \\
  A^\ast k_- & 0 & \frac{\ome_d^2}{c^2} & 0 \\
  0 & A k_+ & 0 & \frac{\ome_d^2}{c^2} \\
\end{array}\right) ,   
\ee
where $k_{\pm}=k_x\pm i k_y$, and $A$ is the coupling coefficient. The
double Dirac-cone appears at the situations with $p$-$d$ degeneracy,
$\ome_p=\ome_d\equiv \ome_0$. The group velocity for the double
Dirac-cone dispersion at $\Gamma$ point is then $\pm
\frac{|A|c^2}{2\ome_0}$ (positive group velocity for bands above the
Dirac point, negative group velocity for bands 
below). The $p$ bands behave as the valence band and the $d$ bands
behave as the conduction band in our photonic crystals. Note that the
coupling between the $p$ and $d$ bands are within the same
pseudo-spin, i.e., between $p_+$ and $d_+$, or between $p_-$ and
$d_-$. The Berry phase for a loop circulating the Dirac point is $\pm
\pi$ for spin up/down bands above the Dirac point. The total Berry
phase is zero, in accordance with time-reversal symmetry.  
\end{widetext}

The physics described by Eq.~(5) resembles that of the quantum spin
Hall effect in electronic systems. The $p$-$d$ inversion at the
$\Gamma$ point leads to the formation of photonic $Z_2$ topological
insulators which have helical edge states. The phase transition from
normal photonic band gaps with trivial topology to the photonic $Z_2$
topological insulator takes place at the black line in Fig.~2, where
the double Dirac cone emerges. The key information in Fig.~2 is the
appearance of two regions support photonic topological insulators (the
two blue regions). This takes place for hollow dielectric cylinders
with large outer and inner radii, or for hollow air cylinders with
small inner radius. We remark that although the $p$-$d$ band gap of
the $Z_2$ topological insulator can be quite large at the $\Gamma$
point, $\sim 40\%$, for both normal and reversed structures, the complete
photonic band gap is reduced, particularly for the inverse
structure. The photonic $Z_2$ topological insulators have helical edge
states which can enable unprecedented manipulation of light flow. For
example, light propagation can be controlled by the orbital angular
momentum. We shall discuss the properties of the edge states below.

\begin{figure}
\begin{center}
\includegraphics[width=7cm]{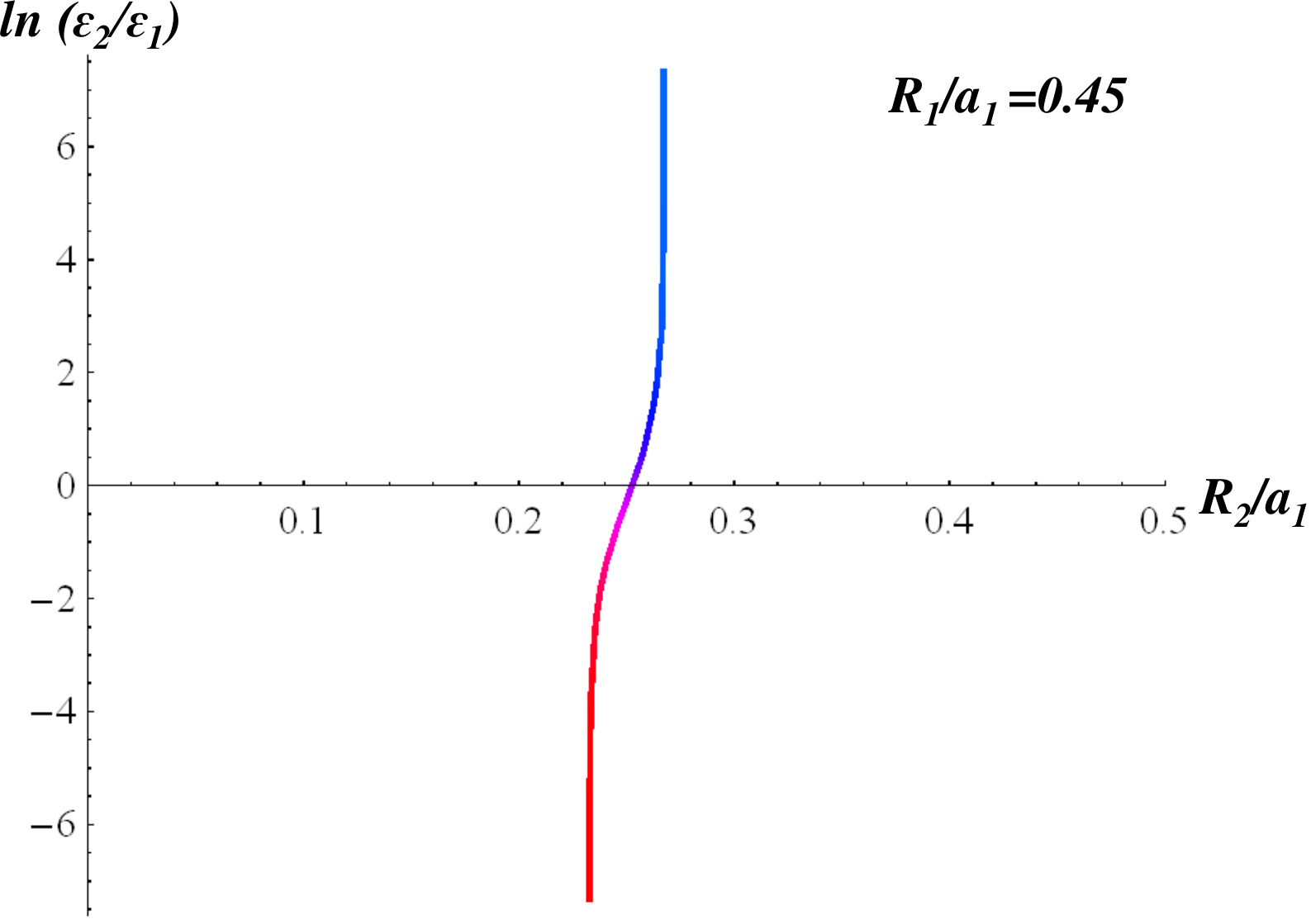}
\caption{ (Color online) Relationship between $\ln(\vep_1/\vep_2)$ and the
  critical inner radius $R_2$ where the double Dirac cone emerges. The
  outer radius of the hollow cylinder is fixed at $R_1 =0.45$. }
\end{center}
\end{figure}

It is natural to ask how the phase diagram changes when the
permittivity ratio $\vep_1/\vep_2$ is tuned. In most photonic crystals
the photonic band gap increases with the permittivity ratio
$\vep_1/\vep_2$\cite{book}. Here we find that, quite interestingly,
the phase boundary between the $Z_2$ topological insulator and the
normal photonic band gap changes negligibly for a {\em very broad
  range} of the permittivity ratio. This interesting property is
demonstrated in Fig.~4 where we examine the phase boundary along the
$R_2$ axis for different permittivity ratio $\vep_1/\vep_2$ for fixed
outer radius $R_1$. The critical value of $R_2$ where the double Dirac
cone emerges is insensitive to the permittivity ratio $\vep_1/\vep_2$
in a very wide range. Although the calculation is done for $R_1=0.45$,
the observed behavior holds true for other values of the outer radius
$R_1$. In fact, we have chosen the $R_1$ in the calculation such that
the dependence of the critical value of $R_2$ on the permittivity
ratio $\vep_1/\vep_2$ is the strongest. The regions with negative
value of $\ln(\vep_1/\vep_2)$ in Fig.~4 stands for the inverse
structure photonic crystal. The horizontal axis represents the case
with $\vep_1/\vep_2=1$, i.e., the homogenous limit. In the homogeneous
limit the photonic band gaps vanish and all the photonic bands become
plane waves. Thus they cannot be associated with the $s, p, d,$ and
$f$ symmetries. The homogeneous limit is a singular limit for the
discussion the topological nodal points and $p$-$d$ inversion. This
explains the discontinuity of, e.g., the $s$-$p$ band 
crossing curve for the normal and inverse structure in the phase
diagram upon crossing the homogeneous limit. Nonetheless, the $p$-$d$
band crossing seems to undergo a ``continuous transition'' from the
normal structure to the inverse structure. The possible physical
scenario in approaching the homogeneous limit is that the plane-wave
component continuously increase to 100\%, while the $s, p, d, f$ wave (the
local Mie resonances of the hollow cylinder) components gradually
vanish in approaching the homogeneous limit.

\begin{widetext}

\begin{figure}
\begin{center}
\includegraphics[width=14cm]{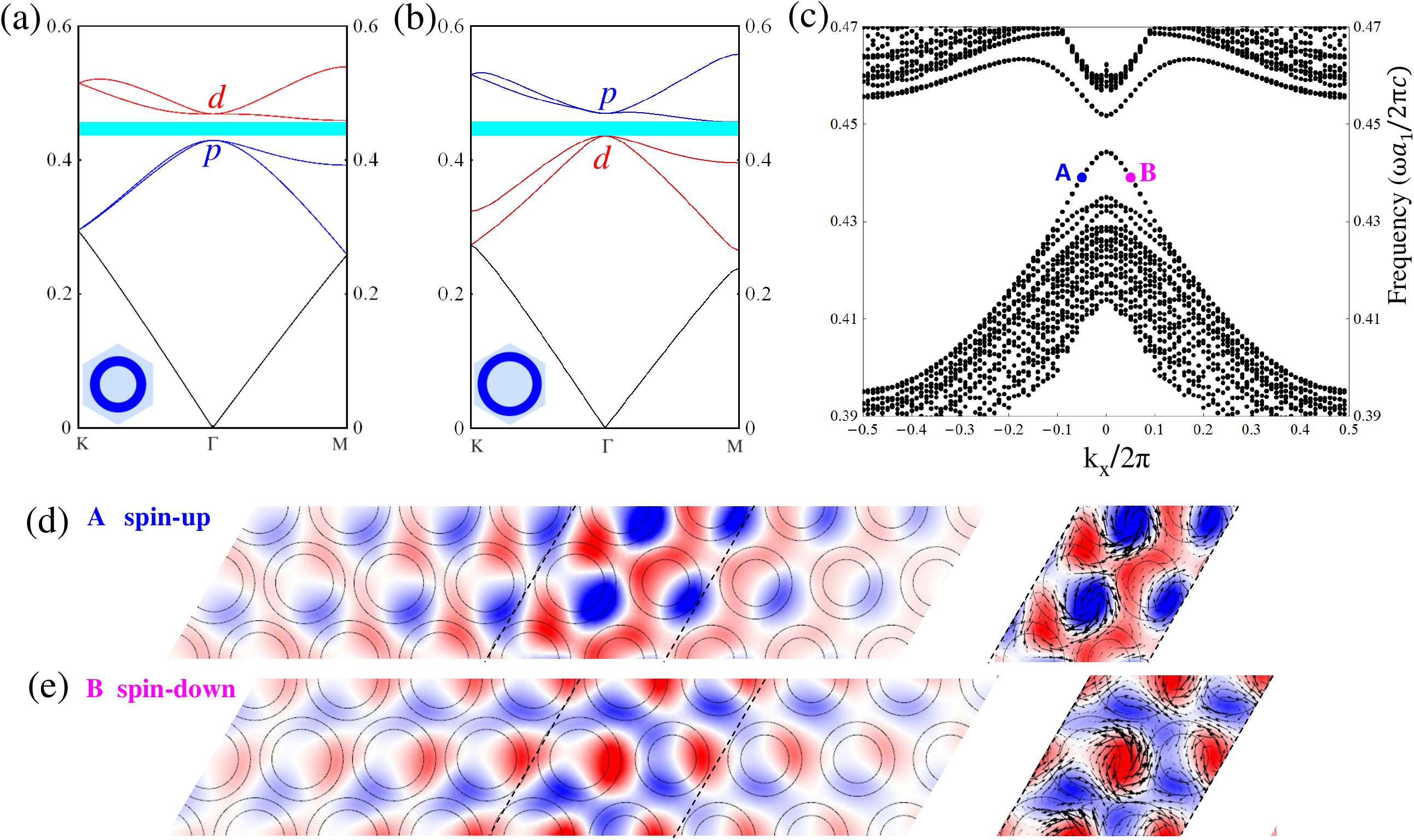}
\caption{ (Color online) Topology induced edge states. (a) The normal
  band structure of the $G$ point in phase diagram with $R_1$=0.40 and
  $R_2$=0.26. (b) The $p$-$d$ reversed band structure of the $H$ point
  in the phase diagram ($R_1=0.45$ and $R_2=0.32$). The common
  complete band gap is marked with the cyan ribbon. (c) The projected
  band structure of two PhCs with oblique line edge. $A$ and $B$ mostly
  comprise of the pseudo-spin-up and spin-down edge states,
  respectively. (d) and (e) are the $E_z$ field pattern of $A$ and $B$,
  respectively. The time-averaged Poynting vectors $\vec{S}=\Re[{\vec
    E}\times{\vec H}^\ast]/2$ near the boundary (between the two
  dashed lines) are shown by the  black arrows.}
\end{center}
\end{figure}

\begin{figure}
\begin{center}
\includegraphics[width=14cm]{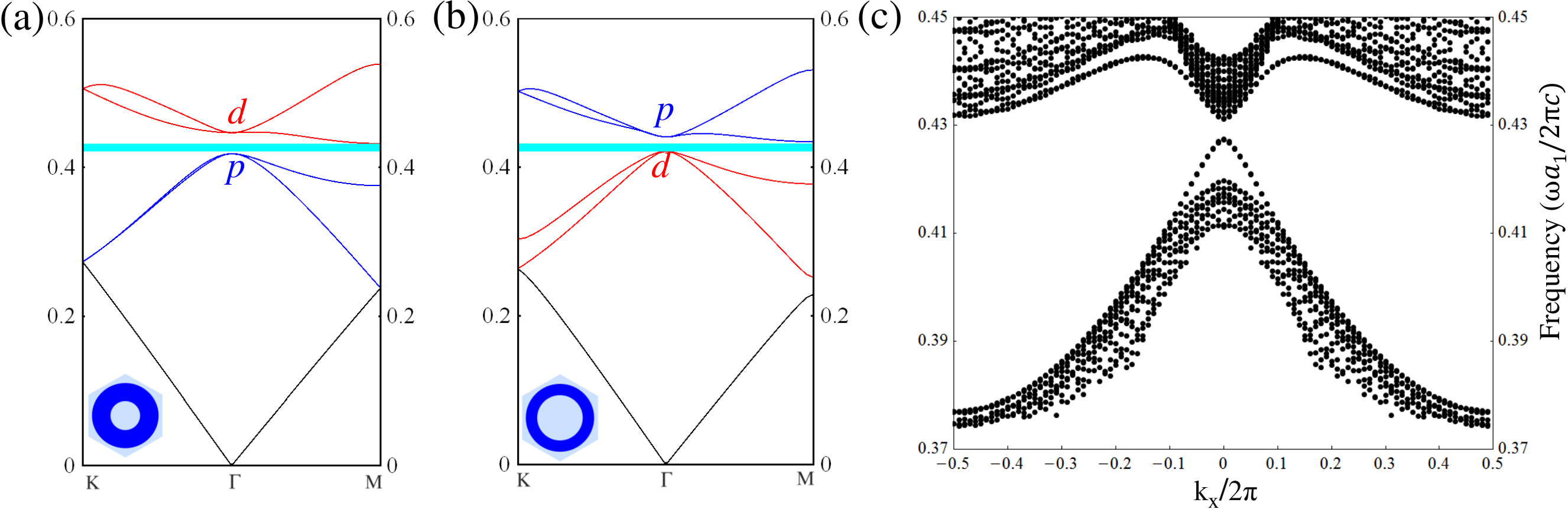}
\caption{ (Color online) Photonic band structure and topological edge
  states. (a) The normal band structure with $R_1=0.45$, $R_2=0.2$, $\vep_1=1$,
  and $\vep_2=9$. (b) The $p$-$d$ reversed band structure with $R_1=0.45$, $R_2=0.3$,
  $\vep_1=1$, and $\vep_2=12$. They have a common band gap marked with cyan
  ribbon. (c) Projected band structure of the two PhCs with a line
  boundary. 
}
\end{center}
\end{figure}

\end{widetext}

\section{EDGE STATES OF THE $Z_2$ TOPOLOGICAL INSULATORS}

We now discuss the properties of the edge states of the photonic $Z_2$
topological insulators. First, let us consider the $G$ and $H$ points in
Fig.2. The $G$ point has normal band structure, while the $H$ point is a
photonic $Z_2$ topological insulator. There is a common complete band gap
marked by the cyan region in Figs.~5(a) and 5(b). When these two photonic crystals
are put together, there is topology induced edge states. The $Z_2$
topology is protected by the pseudo-time-reversal symmetry. The
pseudo-time-reversal operation is ${\cal T}_p=i\hat{\sigma}_y{\cal T}$
with ${\cal T}=\hat{K}$ (the complex conjugation operator) being the
genuine time-reversal operator. Here $\hat{\sigma}_y$ is an operator
acting on the pseudo-spin space for both the $p$ and $d$ bands which
can be written as a combination of the $C_{6v}$ symmetry
operations (see Ref.~\onlinecite{huxiao}) to ensure that ${\cal
  T}_p^2=-1$. The latter 
is crucial in mimicking electronic $Z_2$ topological insulator in
photonic crystals. The pseudo-spin matrices transform as 
follows, under pseudo-time-reversal operation, $\sigma_i\to -
\sigma_i$ for $i=x,y,z$. Under the genuine time-reversal operation,
$\sigma_z\to -\sigma_z$, $\sigma_x\to \sigma_x$, and
$\sigma_y\to\sigma_y$, because $\hat{K}|\up\rangle=|\down\rangle$ and
$\hat{K}|\up\rangle=|\down\rangle$ (since $|\up\rangle=|p_x+ip_y\rangle$ and
$|\down\rangle=|p_x-ip_y\rangle$ for $p$ bands, while
$\up\rangle=|d_{x^2-y^2}+id_{xy}\rangle$ and
$|\down\rangle=|d_{x^2-y^2}-id_{xy}\rangle$ for the $d$ bands).

The $C_{6v}$ symmetry breaks down at the boundary between
the two photonic crystals. Hence the boundary introduces perturbations
that gap the helical edge states. The size of the gap depends on the
strength of the perturbation. However, since the bulk topology of the
two photonic crystals is distinct, there is topology guaranteed
edge states, although they may {\em not} be gapless. This is the 
essential feature for bosonic analog of the $Z_2$ topological
insulators. A typical calculation of the edge states is presented in
Fig.5(c). The helical edge states are gapped at $k_x=0$. Away from the
$k_x=0$ point the helical feature of the edge states is clearly
demonstrated in Figs.5(d) and 5(e). The edge states at the $B$ point is
mostly spin-down (due to the perturbation that gaps the $k_x=0$ point,
the $B$ point  mix slightly with spin-up) as recognized from the
real space distribution of the Poynting vector. Similarly the $A$ point
is mostly spin up and it has positive group velocity, while the $B$
point has negative group velocity. Hence the helical character of the
edge states is preserved except near the $k_x=0$ point.

In some situations the two photonic crystals do not share a common
photonic band gap. We can slightly change the $\vep_1$ or $\vep_2$ of the
structure to get the common photonic band gap. Since the $p$-$d$ inversion
depends negligibly on the ratio of the two permittivity, this does not
change the topological properties of the photonic crystals. For
example, if the photonic crystal with normal band structure has
$R_1$=0.45, $R_2$=0.2, $\vep_1$=9, and $\vep_2$=1 [see in Fig.~6(a)],
while the photonic crystal with nontrivial $Z_2$ topology has
$R_1$=0.45, $R_2$=0.3, $\vep_1$=12, and $\vep_2$=1 [see in
Fig.~6(b)]. The two photonic crystals have a common band gap marked by
the cyan ribbon. When those two photonic crystals are put together the
topology induced edge states are found [see in Fig.~6(c)].

From symmetry considerations, the boundaries break the
pseudo-time-reversal symmetry but still keeps the (genuine)
time-reversal symmetry. The spin-operators that are even under
time-reversal operation are the $\hat{\sigma}_x$ and $\hat{\sigma}_y$.
Thus there can be two types of ``mass terms'' that gap the edge
states. The general form of the edge Hamiltonian that obeys the
time-reversal symmetry is ${\cal H}_{edge}= v
k_x\hat{\sigma}_z + m_x\hat{\sigma}_x + m_y\hat{\sigma}_y$ where $v$ is the group
velocity, $m_x$ and $m_y$ are two real quantities. The magnitude of
the two masses, $m_x$ and $m_y$, depend on the specific
geometry of the boundary. The energy gap at $k_x=0$ is
$2\sqrt{m_x^2+m_y^2}$. Finally, we remark that if $m_x$ and $m_y$ are 
position-dependent, they can induce additional Berry-phase that modulates
photon propagation along the boundary.

\section{APPLICATION POTENTIALS}

The effective refractive index of the photonic crystal can be positive
or negative, depending on the frequency of the light. By matching the
frequency and wave vector parallel to the boundary, we find, in
agreement with Ref.~\onlinecite{refrac}, that the effective refractive index
is frequency dependent, 
\be
n(\ome) =
\frac{\sin(\theta_1)}{\sin(\theta_2)}=\frac{2\ome_0}{|A|c}\left(\frac{\ome-\ome_0}{\ome}\right) ,
\ee
where $\theta_1$ and $\theta_2$ are the angle of incidence and
refraction, respectively. $\ome_0$ is the frequency of the Dirac
point. The above equation demonstrates that the effective refractive
index can be tuned via the frequency. Both positive and negative
refractive indices can be achieved, as indicated in Fig.~7. We
emphasize that the effective refractive index here are in the range
range $-1<n(\ome)<1$, which are unattainable for natural (lossless)
materials. Therefore, the photonic crystal with double Dirac cone can
serve as a particular type of lossless metamaterial with unprecedented
ability of manipulating light.

\begin{figure}
\begin{center}
\includegraphics[width=8.6cm]{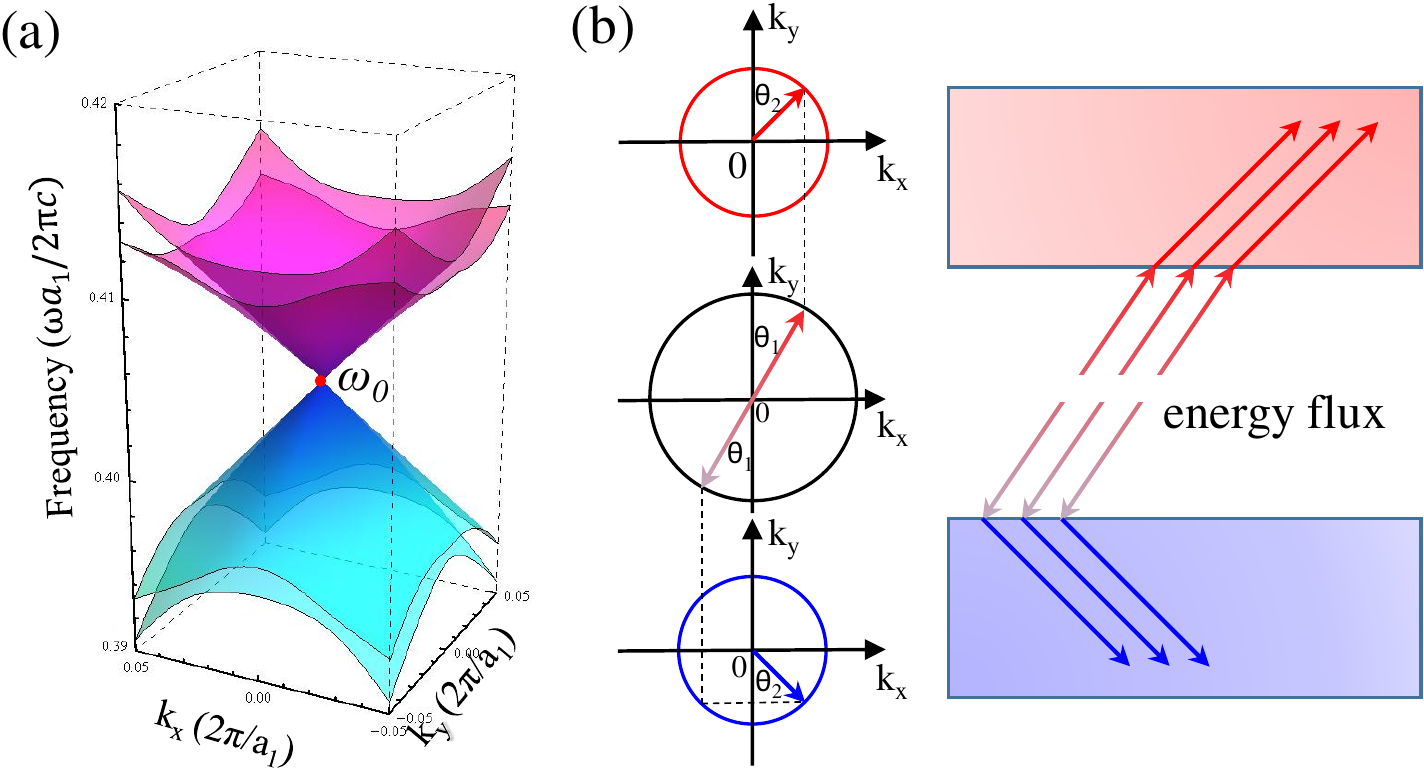}
\caption{ (Color online) Optical property near double Dirac Cone. (a)
  The band structure near double Dirac cone with $R_1=0.45$ and
  $R_2=0.2656$. Four cone-like surfaces touch at $\ome_0$ of the
  $\Gamma$ point. (b) The iso-frequency surfaces for the frequency of
  higher (in red) and lower (in blue) than $\ome_0$, respectively. In
  the middle it is the iso-frequency surface of light in
  air. The refraction law is derived from the conservation of
  frequency and the wave vector parallel to the interface, kx. Right
  panel: the property of positive refraction (upper plot) and negative
  refraction (lower plot).
}
\end{center}
\end{figure}

The edge states can be viewed as unconventional waveguides that
correlates the direction of light flow and its group velocity with the
angular momentum of light (as demonstrated in Fig.5). In addition,
when the frequency of the light is in the bulk photonic bands, the
propagation of light are influenced by the Berry phase in the bulk
bands due to the pseudo-spin-orbit coupling, which leads to the spin
Hall effect of light\cite{berry,onoda2}. The spin-dependent Berry
phase can lead to anomalous velocity and shift in a
scattering/reflection set-up\cite{onoda2}. Due to the time-reversal
symmetry, these anomalous velocity and shift are opposite for opposite
pseudo-spins. These properties can be exploited as angular-momentum
selective transmission and filtering for advanced photonic circuits.

\section{CONCLUSION AND DISCUSSIONS}
Using a simple architecture of core-shell triangle photonic crystal
with $C_{6v}$ point group symmetry we have systematically studied the
evolution of the nodal points in photonic energy bands for various
geometry and (isotropic) permittivity parameters. We show that such a
simple photonic crystal can support $Z_2$ photonic topological
insulators. Compared with previous proposal of $Z_2$ photonic topological
insulators, our architecture is much simpler and hence feasible for
fabrications and experiments. The edge states of the $Z_2$ topological
insulators, unlike for electronic systems, are not ensured to be
gapless. Nevertheless, they show properties similar to the helical
edge states in electronic systems, such as spin-dependent propagation
directions. We also give a full phase diagram for the topological
nodal points, Dirac-like cones and double Dirac cones, for various
geometric parameters. These topological nodal points are proximate to
topologically nontrivial photonic band gaps. The physical origin of
the topological nodal points as well as their properties such as Berry
phase, (pseudo-)spin-orbit coupling are revealed. Dirac cones at
${\vec k}=0$ can be exploited to tune the effective refractive index
to the range $-1<n(\ome)<1$ which are unattainable in 
natural (lossless) materials. These properties, together with the 
Berry phases of photon, offer great potential for future advanced
photonics.


\section*{Acknowledgements} 
We thank supports from the National Science Foundation of China for
Excellent Young Scientists (grant no. 61322504). J.H.J acknowledges
supports from the faculty start-up funding of Soochow University. He
also thanks Sajeev John, Xiao Hu, and Zhi Hong Hang for helpful discussions.

{}


\begin{thebibliography}{999}

\bibitem{GL} V.L. Ginzburg and L.D. Landau, Zh. Eksp. Teor. Fiz. {\bf 20},
  1064 (1950). 

\bibitem{qhe} K. v. Klitzing, G. Dorda, and M. Pepper,
  Phys. Rev. Lett. {\bf 45}, 494 (1980).

\bibitem{fqhe} D. C. Tsui, H. L. Stormer, and A. C. Gossard,
  Phys. Rev. Lett. {\bf 48}, 1559 (1982).

\bibitem{graphene} A. H. Castro Neto, F. Guinea, N. M. R. Peres,
  K. S. Novoselov, and A. K. Geim, Rev. Mod. Phys. {\bf 81}, 109 (2009).

\bibitem{TIreview} M. Z. Hasan and C. L. Kane, Rev. Mod. Phys. {\bf 82},
  3045 (2010); X.-L. Qi and S.-C. Zhang, Rev. Mod. Phys. {\bf 83}, 1057 (2011).

\bibitem{Raghu} F. D. M. Haldane and S. Raghu, Phys. Rev. Lett. {\bf 100},
  013904 (2008); S. Raghu and F. D. M. Haldane, Phys. Rev. A {\bf 78}, 033834 (2008). 

\bibitem{phTI} Z. Wang, Y. Chong, J. D. Joannopoulos,
  M. Solja\v{c}i\'{c}, Nature (London) {\bf 461}, 772 (2009);
  M. C. Rechtsman, J. M. Zeuner, Y. Plotnik, Y. Lumer, D. Podolsky,
  F. Dreisow, S. Nolte, M. Segev, and A. Szameit, Nature (London) {\bf
    496}, 196 (2013); L. Lu, J. D. Joannopoulos, M. Solja\v{c}i\'{c},
  Nat. Photon. {\bf 8}, 821 (2014); Z.-G. Chen, X. Ni, Y. Wu, C. He, X.-C. Sun, L-Y. Zheng,
  M.-H. Lu, and Y.-F. Chen, Sci. Rep. {\bf 4}, 4613 (2014).

\bibitem{eli} E. Yablonovitch, Phys. Rev. Lett. {\bf 58}, 2059 (1987).

\bibitem{sajeev} S. John, Phys. Rev. Lett. {\bf 58}, 2486 (1987).

\bibitem{book} J. D. Joannopoulos, S. G. Johnson, J. N. Winn, and
  R. D. Meade, {\it Photonic Crystals: Molding the Flow of Light} (Princeton
  University Press, New Jersey, 2008).

 \bibitem{qahe} C.-X. Liu, X.-L. Qi, X. Dai, Z. Fang,
   S.-C. Zhang, Phys. Rev. Lett. {\bf 101}, 146802 (2008).

\bibitem{ph-gra} M. C. Rechtsman, Y. Plotnik, J. M. Zeuner, D. Song,
  Z. Chen, A. Szameit, and M. Segev, Phys. Rev. Lett. {\bf 111}, 103901 (2013).


\bibitem{ct2d} X. Huang, Y. Lai, Z. H. Hang, H. Zheng, and C. T. Chan,
  Nat. Mater. {\bf 10}, 582 (2011).

\bibitem{Sakoda} K. Sakoda, Opt. Express {\bf 20}, 3898 (2012); {\bf
    20}, 9925 (2012).


\bibitem{z2-ph} A. B. Khanikaev, S. H. Mousavi, W. K. Tse,
  M. Kargarian, A. H. MacDonald, and G. Shvets, Nat. Mater. {\bf 12}, 233
  (2013).

\bibitem{huxiao} L. H. Wu and X. Hu, Phys. Rev. Lett. {\bf 114}, 223901 (2015).

\bibitem{pgs} G. van Miert and C. M. Smith, Phys. Rev. B {\bf 93},
  035401 (2016).


\bibitem{mei} Y. Li, Y. Wu, and J. Mei. Appl. Phys. Lett. {\bf 105},
  014107 (2014).



\bibitem{colloid} K. P. Velikov, A. Moroz, and A. van Blaaderen,
  Appl. Phys. Lett. {\bf 80}, 49 (2002).

\bibitem{bio} A. R. Parker, R. C. McPhedran,
D. R. McKenzie, L. C. Botten, and N.-A. P. Nicorovici, Nature (London)
{\bf 409}, 36 (2001).


\bibitem{sa-book} K. Sakoda, Optical Properties of Photonic Crystals,
  2nd ed. (Springer, Berlin, 2005).



\bibitem{tb} E. Lidorikis, M. M. Sigalas, E. N. Economou, and
  C. M. Soukoulis, Phys. Rev. Lett. {\bf 81}, 1405 (1998)


\bibitem{kivshar} M. V. Rybin, D. S.Filonov, K. B. Samusev, P. A. Belov, Y. S. Kivshar, and M. F. Limonov, Nat. 
Commun. {\bf 6}, 10102 (2015).


\bibitem{fuprb} L. Fu and C. L. Kane, Phys. Rev. B {\bf 76}, 045302 (2007).


\bibitem{onoda} T. Ochiai and M. Onoda, Phys. Rev. B {\bf 80}, 155103 (2009).

\bibitem{mei2} Y. Li, Y. Wu, X. Chen, and J. Mei. Opt. Express {\bf 21}, 7699 (2013).

\bibitem{mei3} J. Mei, Y. Wu, C. T. Chan, and Z.-Q. Zhang,
  Phys. Rev. B {\bf 86}, 035141 (2012).

\bibitem{refrac} L. Wang, S.-K. Jian, and H. Yao, arXiv:1511.09282.

\bibitem{berry} M. Barkeshli and X.-L. Qi, Phys. Rev. Lett. {\bf 107},
  206602 (2011).

\bibitem{onoda2} M. Onoda, S. Murakami, and N. Nagaosa,
  Phys. Rev. Lett. {\bf 93}, 083901 (2004); Phys. Rev. E {\bf 74}, 066610 (2006).


\end{thebibliography}
\end{document}